\documentclass[reprint,aps,prb,twocolumn,showpacs,superscriptaddress,longbibliography]{revtex4-2}

\usepackage{graphicx}  
\usepackage{float}
\usepackage{subfigure}
\usepackage{multirow}
\usepackage{tikz}
\usepackage{fancyhdr}
\usepackage{longtable}
\usepackage{parskip}
\usepackage[T1]{fontenc}
\usepackage{dcolumn}   
\usepackage{bm}        
\usepackage{amsfonts}  
\usepackage{amsmath}   
\usepackage{amssymb}   

\newcommand{\brac}[1]{\left[ #1 \right]}

\newcommand{\pwisein}{\left\{ \begin{array}{ll}}
\newcommand{\pwiseout}{\end{array}\right.}

\usepackage[normalem]{ulem}
\usepackage[english]{babel}
\begin{document}
\title{Comparative study of magnetic properties of Mn$^{3+}$ magnetic clusters in GaN using classical and quantum mechanical approach}

\author{Y.~K.~Edathumkandy} 
\affiliation{Institute of Physics, Polish Academy of Sciences, PL 02-668 Warszawa, Poland}

\author{D.~Sztenkiel} \email{sztenkiel@ifpan.edu.pl}
\affiliation{Institute of Physics, Polish Academy of Sciences, PL 02-668 Warszawa, Poland}


\date{\today}

\begin{abstract}  
Currently, simulations of many-body quantum systems are known to be computationally too demanding to be solved on classical computers. The main problem is that the computation time and memory necessary for performing the calculations usually grow exponentially with the number of particles $N$. An efficient approach to simulate many-body quantum systems is the use of classical approximation. However, it is known that at least at low temperatures, the allowed spin fluctuations in this approach are overestimated what results in enhanced thermal fluctuations. It is therefore timely and important to assess the validity of the classical approximation. To this end, in this work, we compare the results of numerical calculations of small Mn$^{3+}$ magnetic clusters in GaN, where the Mn spins are treated classically with those where they are treated quantum-mechanically (crystal field model). In the first case, we solve the Landau-Lifshitz-Gilbert (LLG) equation that describes the precessional dynamics of spins represented by classical vectors.
On the other hand, in the crystal field model, the state of Mn$^{3+}$ ion ($d^4$ configuration with $S=2$, $L=2$) is characterized by the set of orbital and spin quantum numbers $|m_s,m_L>$. Particular attention is paid to use numerical parameters that ensure the same single ion magnetic anisotropy in both classical and quantum approximation. Finally, a detailed comparative study of magnetization $\mathbf{M}(\mathbf{H},T)$ as a function of the magnetic field $\mathbf{H}$, temperature $T$, number of ions in a given cluster $N$ and the strength of super-exchange interaction $J$, obtained from both approaches will be presented.


\end{abstract}


\maketitle 

\section{Introduction}
\begin{figure}[htp]
    \centering
    \includegraphics[width=8cm]{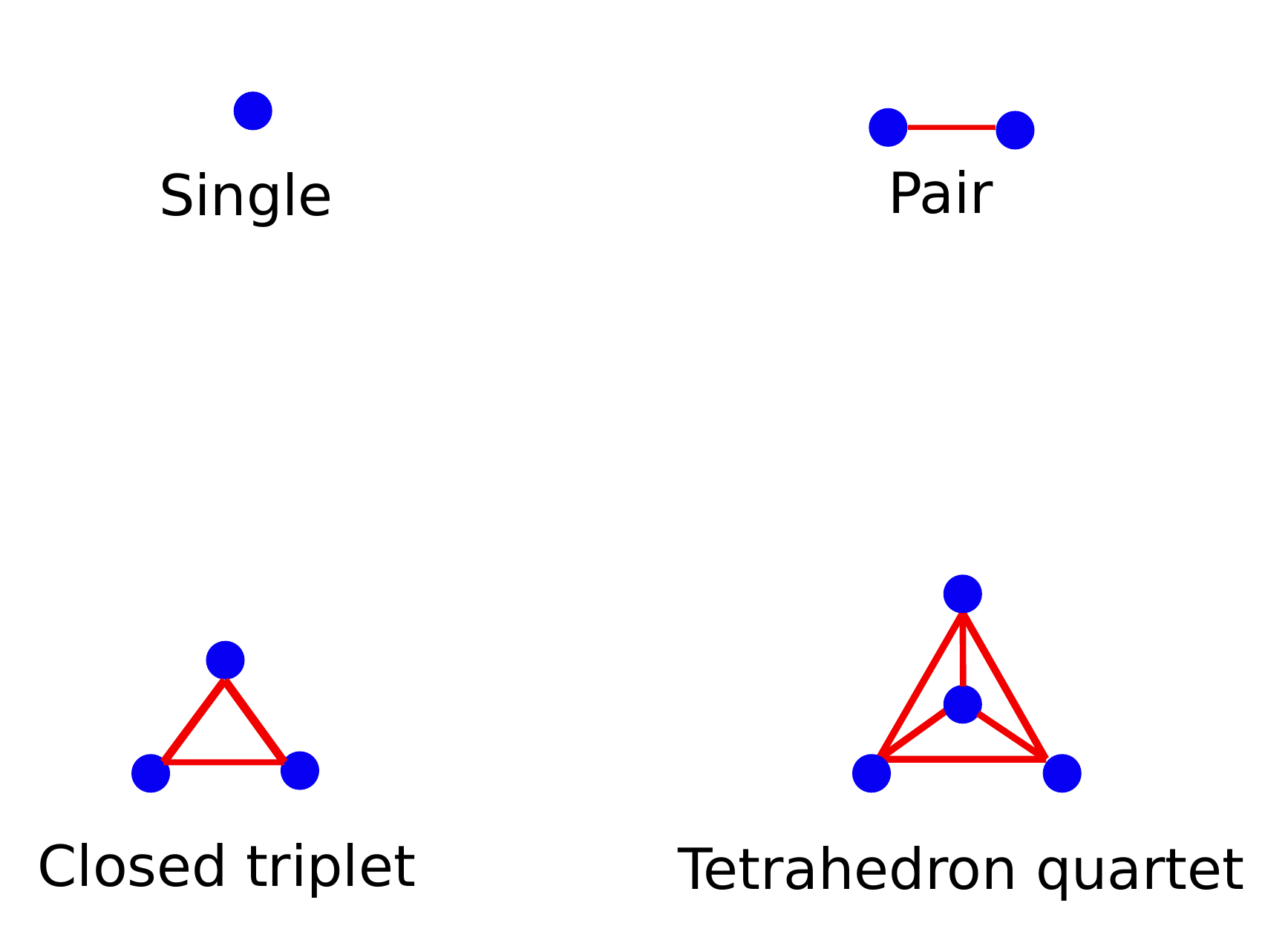}
    \caption{Selected clusters composed of magnetic ions with a single exchange constant. Lines indicate exchange bonds. Only clusters with up to four ions are considered.}
    \label{fig:galaxy}
\end{figure}

Simulations of quantum systems using classical computers bear the potential for advancing scientific knowledge of properties of many materials and their time evolution. However, such simulations may cause diverse problems. Difficulties arise from the exponential growth of required resources (computing time and computer memory) with the number of elements in the system \cite{e12112268,PhysRevX.10.041038, meyer2002quantum}. For example, the number of parameters required to describe the state of $N$ interacting fermionic particles scales as $2^{N}$. Consequently, the memory allotted to save the state increases exponentially with $N$ necessitating the use of vast gigabytes of data. Similarly, the computing time required for performing the simulations is also subject to the same exponential scaling with the number of particles, demanding, in some cases, several years to complete the simulation. Even though approximation techniques such as quantum Monte-Carlo methods \cite{foulkes2001quantum} and renormalisation group algorithms developed by Wilson \cite{wilson1971renormalization} have been designed, they are not entirely successful in the sufficient description of the state of the systems under investigation. Fortunately, many-body quantum systems can be, in principle, simulated using classical approximations. The Landau-Lifshitz-Gilbert (LLG) equation with Langevin dynamics has been applied extensively to compute magnetic properties of different materials and nanoparticles. With the significant improvement of the computing power and efficiency of parallel algorithms, large-scale atomistic spin model or micromagnetic simulations using the stochastic LLG (sLLG) method gradually become possible. For example, ferromagnetic hysteresis loops, spin waves, domain wall motion, and precessional or thermally assisted magnetization switching processes have been investigated using this approach \cite{evans2014atomistic, evans2015quantitative, Schlickeiser:2014_PRL, Ellis:2015_APL, Bender:2017_PRL, Hinzke:2011_PRL}. Classical approximations can reduce the overall computational complexity, but they usually face difficulties in proper reproduction of temperature-dependent properties of the system due to neglecting of spin quantization. It is generally accepted that allowed spin fluctuations in the classical approach are overestimated and an appropriate re-scaling of simulation temperature is required to resolve this issue \cite{evans2014atomistic, evans2015quantitative}. It is important then to somehow assess the validity of the classical approximation and its underlying assumptions. To accomplish this goal, we compare magnetizations $\mathbf{M}(\mathbf{H},T)$ as a function of the magnetic field ($\mathbf{H}$) and temperature ($T$) obtained by the sLLG method with those obtained by the exact approach, that is by the quantum mechanical crystal field model (CFM). Here we choose Ga$_{1-x}$Mn$_x$N dilute magnetic semiconductor, as this material has been extensively and thoroughly investigated \cite{gosk2005magnetic, wolos2004neutral,stefanowicz2010structural, Bonanni:2011_PRB, sztenkiel2016stretching, Gas:2018_JALCOM, Gas:2021_JALCOM, sztenkiel2020crystal,Grodzicki:2019_SS, Stefanowicz:2013_PRB, Sarigiannidou:2006_PRB}, and the experimental findings has been remarkably well reproduced by the exact CFM approach \cite{wolos2004neutral, stefanowicz2010structural, sztenkiel2016stretching, sztenkiel2020crystal}. Even though, there are many different cluster types in GaN with randomly distributed Mn ions \cite{shapira2002magnetization}, in this article we consider only a few exemplary ones. In particular, the simulations are restricted here to very small magnetic clusters consisting of up to four Mn$^{3+}$ ions coupled by the ferromagnetic super-exchange interaction $-J\mathbf{S}_1\cdot\mathbf{S}_2$, as shown in Fig.~\ref{fig:galaxy}.

In this paper, we show that the comparison between classical and quantum approaches improves with the increasing magnitude  of the exchange coupling parameter $J^q$ and it is the best for large number of ions $N$ and strong magnitude of the exchange energy as compared to the thermal one, that is $J^q |\mathbf{\hat{S}}_i| \cdot |\mathbf{\hat{S}}_j| \gg k_B T$. It is shown, that it is possible to only partially explain the inconsistency between classical and quantum results presented in temperature-dependent magnetization curves by referring to the concept of quantum mechanical “stiffness”. On the contrary, we find that the key factor is the value of effective spin of the system. Depending on the relative strength between the thermal and the exchange energy we deal with uncoupled ions characterized by $S=2$ or strongly bound clusters with large total spin value $S_T = N S$. For weak coupling between the spins $J^q |\mathbf{\hat{S}}_i| \cdot |\mathbf{\hat{S}}_j| < k_B T$, the investigated clusters consist of practically uncoupled spins. Then the classical approximation underestimates the actual value of magnetization. Contrary, in the strong coupling regime, where $J^q |\mathbf{\hat{S}}_i| \cdot |\mathbf{\hat{S}}_j| \gg k_B T$, we deal with one magnetic system characterized by high total spin value $S_T = NS$. In this limit, $S_T$ increases with the number of ions $N$ in the cluster, which improves the comparison between two approaches. These conclusions are consistent with the common relation, that the classical and quantum results differ for a small value of $S$ and both treatments merge for $S \rightarrow \infty$.

\section{MODEL}
\subsection{Crystal field model}

 When an ion is inserted in a solid, Coulomb interaction of its charge distribution with the neighbouring charges in the crystal together with the hybridization of relevant atomic orbitals should be taken into account. This effect is called the crystal field interaction, which causes the quenching of the orbital angular momentum and the development of the single-ion magnetic anisotropy. By incorporating in the   Hamiltonian additional terms corresponding to the spin-orbit interaction and the influence of external magnetic field $\mathbf{B}$, it is possible to simulate magnetic properties of transition metal ions placed in a crystal environment. Such simulations, termed crystal field model, were developed by Vallin and coworkers \textit{et al.} to study the magnetization of a Cr ion in II-VI dilute magnetic semiconductors \cite{vallin1974epr}. 
Here we use this model to obtain $\mathbf{M}(\mathbf{H},T)$ curves of magnetic clusters composed of up to 4 Mn ions \cite{sztenkiel2020crystal}. Detailed reviews on this theoretical model are given in Ref.~\onlinecite{vallin1970infrared,vallin1974epr, tracy2005anisotropic, gosk2005magnetic, wolos2004neutral, stefanowicz2010structural, sztenkiel2020crystal} and we follow this approach. The Hamiltonian for an isolated Mn$^{3+}$ impurity in GaN crystal is of the form, 

\begin{equation}
\mathcal{H}^{q}(j)=\mathcal{H}^{q}_{CF}+\mathcal{H}^{q}_{JT}(j)+\mathcal{H}^{q}_{TR}+\mathcal{H}^{q}_{SO}+\mathcal{H}^{q}_{B}
\end{equation}

where $\mathcal{H}^{q}_{CF}=-\frac{2}{3}B_4(\hat{O}_4^0-20\sqrt{2}\hat{O}_4^3)$ is the cubic crystal field of tetrahedral symmetry, $\mathcal{H}^{q}_\textrm{JT}=\tilde{B}_2^0\hat{\Theta}_4^0+\tilde{B}_4^0\hat{\Theta}_4^2$ describes the static Jahn-Teller distortion of tetragonal symmetry and  $\mathcal{H}^{q}_{TR}=B_2^0\hat{O}_4^0+B_4^0\hat{O}_4^2$ represents the trigonal distortion along the GaN hexagonal $\bold{c}$-axis. This model incorporates also the spin orbit coupling $\mathcal{H}^{q}_{S0}=\lambda \hat{\textbf{L}}\cdot\hat{\textbf{S}}$ and interaction with the external magnetic field $\mathcal{H}^{q}_B=\mu_B(g_L\hat{\textbf{L}}+g_S\hat{\textbf{S}})\textbf{B}$ with $g_S=2$, $g_L=1$ and $\mu_B$ representing the Bohr magneton. Here $\hat{\textbf{S}}$ and $\hat{\textbf{L}}$ describe the spin and angular-momentum operator, respectively. The symbols $\hat{\Theta}$ and  $\hat{O}$ represent the equivalent Stevens operators along one of the cubic axis $e_\textrm{JT}^j$ ($j=A,B,C$) and the trigonal axis $\brac{001}\parallel \mathbf{c}$-axis of GaN, respectively. These three vectors $e_\textrm{JT}^j$ produce magnetic easy axes and are given by

\begin{equation} \label{eq1}
\begin{split}
 \scriptstyle \textbf{e}^A_\textrm{JT}=\brac{\sqrt{\frac{2}{3}},0,\sqrt{\frac{1}{3}}},   \textbf{e}^B_\textrm{JT}=\brac{-\sqrt{\frac{1}{6}},-\sqrt{\frac{1}{2}},\sqrt{\frac{1}{3}}}, \\
\scriptstyle \textbf{e}^C_\textrm{JT}=\brac{-\sqrt{\frac{1}{6}},\sqrt{\frac{1}{2}},\sqrt{\frac{1}{3}}}
\end{split}
\end{equation}
In the case of single Mn$^{3+}$ impurity in GaN with 3d$^4$ configuration ($L=2$, $S=2$), the matrix representation of the Hamiltonian is obtained using the uncoupled spin and angular momentum basis states $|m_L,m_S>$ with $m_S=-2,-1,0,+1,+2$ and $m_L=-2,-1,0,+1,+2$. Thus the full matrix has dimensions of ($25$ x $25$). By diagonalization of this matrix, one can obtain the energy eigen-values $E$ and eigen-functions $|\phi>$. Then average magnetic moment of Mn$^{3+}$ ion ($\textbf{M}^q$) can be obtained from the following formula,
\begin{equation}
<\textbf{M}^q>=\frac{1}{Z}(Z_A<\textbf{m}^A>+Z_B<\textbf{m}^B>+Z_C<\textbf{m}^C>)
\end{equation}

where $Z_j=\sum_{i}exp(-E^j_i/k_{B}T)$ is the partition function at the $j$-th JT center, $Z=Z_A+Z_B+Z_C$, and 

\begin{equation}
<\textbf{m}^j>=\frac{-\mu_B\sum_{i=1}^{25}<\phi_i^j| g_L \hat{\textbf{L}}+g_S\hat{\textbf {S}}|\phi_i^j>exp(-E^j_i/k_{B}T)}{\sum_{i=1}^{25}exp(-E^j_i/k_{B}T)}
\end{equation}
The symbols $E^j_i$ and $\phi_i^j$ correspond to $i$-th energy level and eigen-state of Mn$^{3+}$ ion being in the $j$-th JT center.

In this work, we extend the CFM model of single substitutional Mn$^{3+}$ in GaN by considering small clusters of Mn ions coupled by the ferromagnetic super-exchange interaction $\mathcal{H}^q_{Exchange}=-\sum_{ \brac{i,k} } J^q \hat{\textbf{S}}_{i} \cdot \hat{\textbf{S}}_{k}$, where $\brac{i,k}$ denotes all the nearest neighbour pairs within a given cluster and $J^q$ is the value of exchange coupling constant. Since these calculations are computationally demanding, the simulations are restricted to clusters composed of up to four ions. The selected cluster types are shown in Fig.~\ref{fig:galaxy}. Only the ferromagnetic nearest neighbour (nn) interaction is taken into account. Then, the relevant eigen-energy and eigen-functions of a pair, triplet and quartet are found by the numerical diagonalization of full ($25^{2}$x$25^{2}$), ($25^{3}$x$25^{3}$) and ($25^{4}$x$25^{4}$) Hamiltonian matrix respectively (c.f. Ref.~\onlinecite{sztenkiel2020crystal} for details).  Crystal field parameters are listed in Tab.~\ref{table:CFMParameters} and most of their numerical values are taken from Ref.~\onlinecite{sztenkiel2016stretching}.  

\begin{table}[h!]
\centering
\begin{ruledtabular}
\begin{tabular}{ c c c c c c c c } 
 \hline
 \\
 $  B_{4}$  &  $B^0_{2}$  &  $B^0_{4}$  &  $\widetilde B^0_{2}$  &  $\widetilde B^0_{4}$  &    $\lambda_{TT}$  &  $\lambda_{TE}$\\ [1ex] 
 \hline\hline
 \vspace{2mm}
 \\
  11.44  &  3.19  &  -0.423 &  -5.85  &  -1.17  &  5.5  &  11.5  \\ 
 \hline
\end{tabular} 
\end{ruledtabular}
\caption{Parameters of the crystal field  model of Mn$^{3+}$ ion in GaN. All values are in meV.}
\label{table:CFMParameters}
\end{table}

\subsection{Classical model}

The goal of this work is to compare the results of the classical and quantum mechanical approach. Therefore the classical model should mimic the quantum one as closely as possible. Here we use a classical atomic spin model, which treats each atom as possessing a localised magnetic moment. The effective spin Hamiltonian contains contributions from the exchange interactions, magnetocrystalline anisotropies (Jahn-Teller and trigonal deformation)  and interaction with the external magnetic field. The same terms are present in the quantum approach. Now, we neglect the spin-orbit interaction and the cubic crystal field splitting (its contributions to the magnetic anisotropy is negligible).

Therefore, the complete Hamiltonian in the classical approach is given by

\begin{equation}
\mathcal{H}^{c}=\mathcal{H}^{c}_{B}+\mathcal{H}^{c}_{TR}+\mathcal{H}^{c}_{\textrm{JT}}+\mathcal{H}^c_{Exchange}
\end{equation}

where, $\mathcal{H}^{c}_{B}=-\mu_S \sum_{i} \textbf{m}_i \textbf{H}$ is the Zeeman energy describing the interaction between spins and an external magnetic field $\textbf{H}$. Now $\bold{m}$ is local spin moment direction, treated as a classical vector with norm |$\bold{m}$|= 1, and $\mu_S=g\mu_BS$ is the actual atomic moment, with $S=2$. The second and third term represent the magnetic anisotropy energy $\mathcal{H}^{c}_{Anizo}$. Similarly as in the quantum model, $\mathcal{H}^{c}_{Anizo}$ is expressed as a sum of uniaxial $\mathcal{H}_{TR}$ and triaxial $\mathcal{H}_{\textrm{JT}}$ magnetocrystalline anisotropy energy. These terms, in the classical approach, are defined as

\begin{equation}
\mathcal{H}^{c}_{TR}=-\frac{1}{2}K_{TR}\sum_{i}\frac{1}{2}[m^2_{iz}-(m^2_{ix}+m^2_{iy}]
\end{equation}

\begin{equation}
\mathcal{H}^{c}_\textrm{JT}=-\frac{1}{2}K_\textrm{JT}\sum_{i}\sum_{j=A,B,C}(\textbf {m}_i.\textbf {e}^j_\textrm{JT})^4
\end{equation}

$K_{TR}$, $K_{\textrm{JT}}$ are the anisotropy constants for the trigonal and Jahn-Teller distortion, respectively. The last term represents a ferromagnetic super-exchange interaction between spin moments

\begin{equation}
\mathcal{H}^c_{Exchange}=-\sum_{ \brac{i,k} } J^c \textbf{m}_{i} \cdot \textbf{m}_{k}
\end{equation}

The exact value of $J^c$ (and $J^q$) is not known. Ab initio simulations predict that the magnitudes of $J$ are rather high, between 15 \cite{szwacki2011aggregation} and 62~meV \cite{sato2010first}. In our recent work, we approximate the lower limit for the value of nn exchange coupling between Mn ions in GaN, which result in $J \geq 8.6$~meV. All values of $J$ are given assuming that the norm of Mn$^{3+}$ spins $\bold{S}$ are $|\bold{S}|=2$. Here, for comparison purposes, we choose a few different values of the superexchange coupling  parameter $J^q$, that is, 0.25, 0.5, 2 and 10~meV. 

\begin{table}[h!]
\centering
\begin{ruledtabular}
\begin{tabular}{ c c c c c } 
 \hline
 \\
 $  J^{q}$ & 0.25 & 0.5 & 2 &10 \\ [1ex] 
 \hline\hline
 \\
  $  J^{c}$  &  1.5  & 3 & 12  &  60   \\ [1ex] 
 \hline
\end{tabular}
\end{ruledtabular}
\caption{Exchange coupling constants used for quantum ($J^q$) and classical ($J^c$) simulations. All values are in meV.}
\label{table:Jvalues}
\end{table}

One should remember that in classical simulations, according to the standard convention, we deal with magnetic spin moment $\bold{m}$ treated as a classical vector with norm $|\bold{m}|=1$. On the other hand, the norm of the quantized spin operator is $|\bold{\hat{S}}|=\sqrt{S(S+1)}$ what gives $|\bold{\hat{S}}|=\sqrt{6}$ for S=2. Therefore the used values of $J^c$ should be six times larger than those  applied in the crystal field model approach ($J^q$), that is $J^c = S(S+1)J^q$. This ensures the same range of exchange energy in both classical ($-J^c |\bold{m}_1| \cdot |\bold{m}_2$|) and quantum ($-J^q |\bold{\hat{S}}_1| \cdot |\bold{\hat{S}}_2$|) approximations. The values of exchange constants are tabulated in Tab.~\ref{table:Jvalues}.

\begin{figure}[h!]
    \centering
    \includegraphics[width=9 cm]{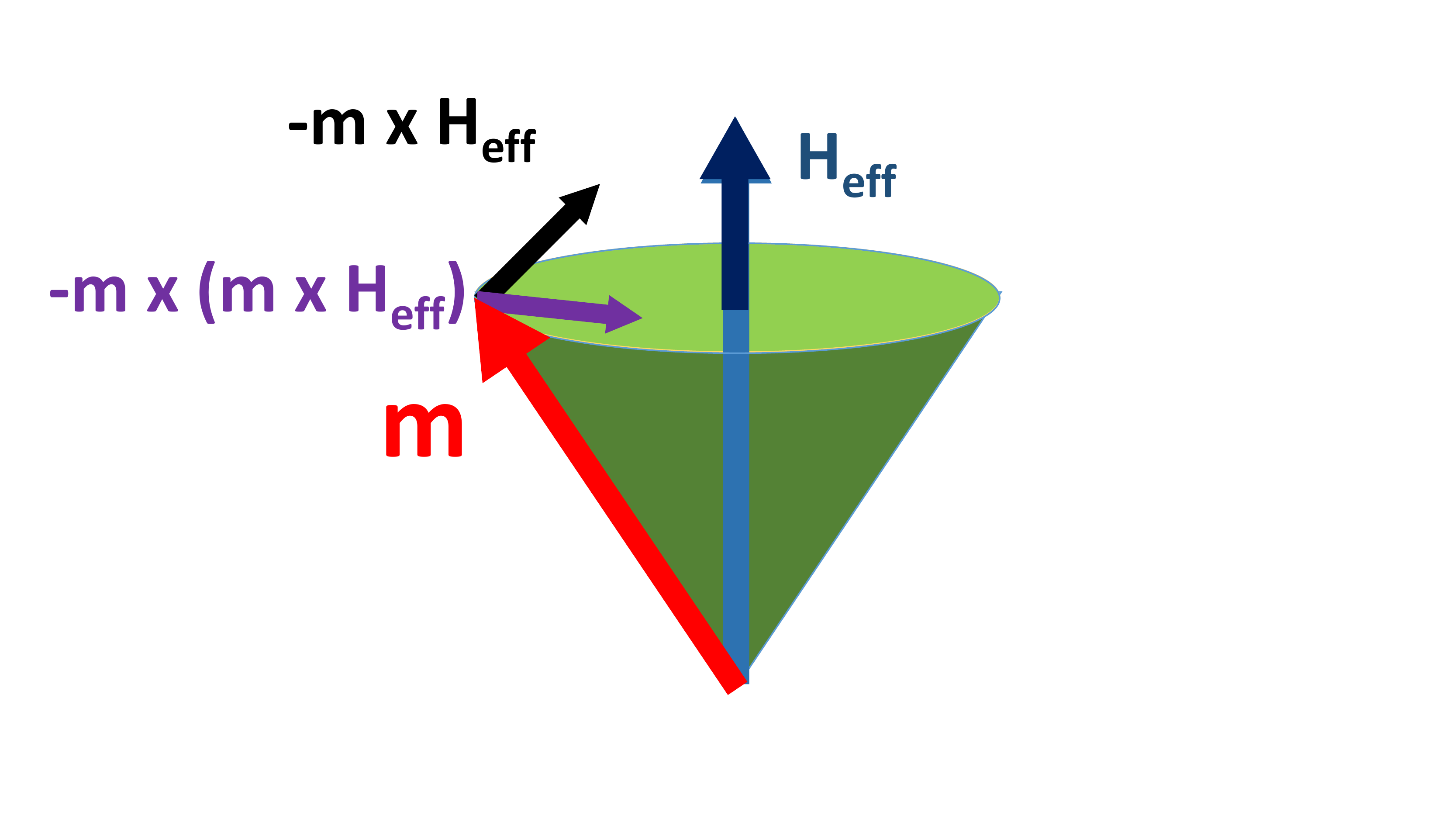}
\caption{(Color online) Motion of a magnetic moment ($\textbf{m}$) in an effective magnetic field ($\textbf{H}_{eff}$). $\textbf{m}\times(\textbf{m}\times\textbf{H}_{eff})$ and $\textbf{m}  \times \textbf{H}_{eff}$ represents damping and precessional terms, respectively.}
    \label{fig:LLG_motion}
\end{figure}

The dynamics of the magnetic spin moment treated as a classical vector in an effective magnetic field $\textbf{H}_{eff}$ is governed by the Landau-Lifshitz-Gilbert equation (Fig.~\ref{fig:LLG_motion}) given by

\begin{equation}
\frac{\partial \textbf{m}_i}{\partial t}=-\frac{\gamma}{1+\alpha ^2}[\textbf{m}_i\times \textbf{H}^i_{eff}+\alpha \textbf{m}_i\times (\textbf{m}_i\times \textbf{H}^i_{eff}]
\end{equation}

where $\alpha=1$ is the Gilbert damping constant, $\gamma$ is the gyromagnetic ratio and the $\textbf{H}^i_{eff}$ is the net effective field acting on the $i-th$ magnetic moment. This effective magnetic field is obtained by calculating the spin Hamiltonian's first derivative.

\begin{equation}
\textbf{H}^i_{eff}=-\frac{1}{\mu_s}\frac{\partial \mathcal{H}}{\partial \textbf{m}_i}
\end{equation}

As we know, the LLG equation is strictly applicable for zero temperature
simulations. Thermal effects cause fluctuations in spin moments, and at elevated temperatures, they can even overwhelm the exchange interactions leading to a ferromagnetic-paramagnetic transition. To investigate the temperature-dependent properties, we use Langevin dynamics by adding a temperature-dependent term to $\textbf{H}_{eff}$. Then, the effective field is written as

\begin{equation}
\textbf{H}^i_{eff}=-\frac{1}{\mu_s}\frac{\partial \mathcal{H}}{\partial \textbf{m}_i}+\textbf{H}^i_{Th}
\end{equation}                                              

The fundamental idea of Langevin dynamics is to represent the thermal fluctuations by a Gaussian white noise term. Width of this Gaussian term increases as the temperature increases, which means a more substantial thermal effect on the system \cite{evans2014atomistic}. The thermal fluctuations are represented by a Gaussian distribution $\bm{\Gamma}(t)$ in 3D space with a mean of zero and a standard deviation of 1. At each time, an instantaneous thermal field acting on each magnetic spin moment is given by 

\begin{equation}
\textbf{H}_{Th}=\bm{\Gamma(t)} \sqrt{ \frac{2 \alpha k_{B} T}{\gamma \mu_{S} \delta t}}
\end{equation}

where $k_{B}$ is the Boltzman constant, $T$ is the simulation temperature, and $\delta t = 0.5\cdot 10^{-5}$~s is the integration time step. Further, we have used $15\cdot 10^{8}$ equilibration and averaging steps for the simulation of our system. The Langevin dynamics with the LLG equation is a stochastic differential equation. One should note that the same results can be obtained using the classical Monte-Carlo method \cite{evans2014atomistic}.\\

\section{Numerical RESULTS}

\subsection{Adjustment of numerical parameters}

\begin{figure}[hbt]
  \centering
  \includegraphics[width=7.6 cm]{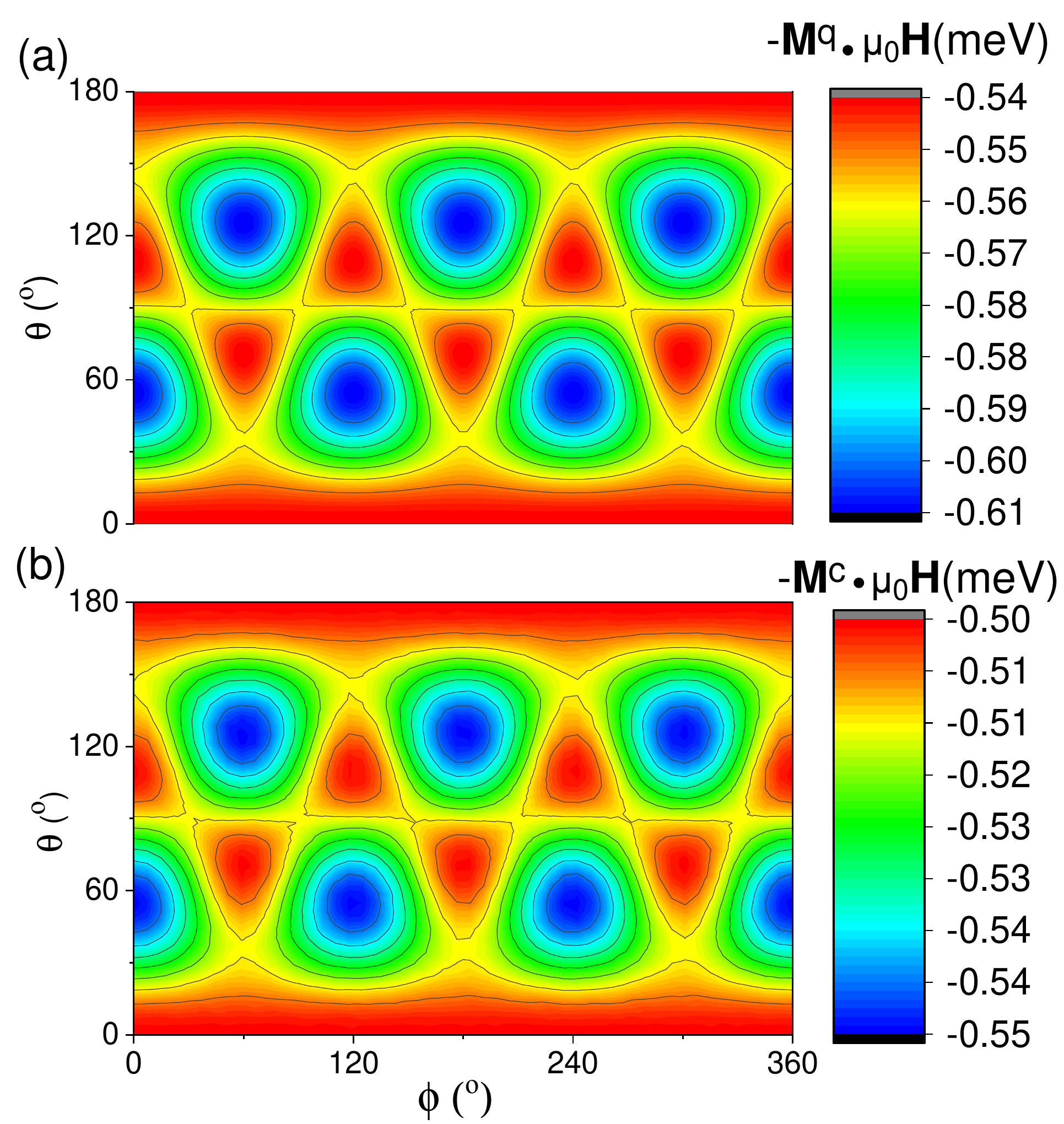}
  \caption{(Color online) The single ion magnetic anisotropy energy $E=-\bold{M} \cdot \mu_0\bold{H}$ at $T=2$~K and $\mu_0H=3$~T, where $\bold{M}=\bold{M}(\bold{H},T)$. $\bold{H}$ can take all possible values on the sphere surface with radius $\mu_0H$. The unit vector describing the direction of given external magnetic field $\bold{H}$ is represented using spherical coordinates $\theta, \phi$, namely $\textbf{e}_H = [\mathrm{sin}\theta \mathrm{cos} \phi,\mathrm{sin}\theta \mathrm{sin} \phi, \mathrm{cos} \theta]$. The magnetization $\bold{M}$ was calculated using (a) the quantum $\bold{M}^q$ and (b) the classical $\bold{M}^c$ approach. The difference in magnetic anisotropy energy for the quantum($\Delta E^q$) and classical ($\Delta E^c$) simulations are 0.07 and 0.05~meV, respectively. No contribution from the trigonal deformation is taken into account.}
  \label{fig:Anisotropy2D}
\end{figure}

In order to perform the comparative study, we should use numerical parameters that ensure the same single ion magnetic anisotropy in both classical and quantum approximation, at least at low temperatures. Crystal field model parameters are listed in Tab.~\ref{table:CFMParameters}. In order to adjust classical anisotropy constants $K_{TR}$ and $K_\textrm{JT}$ we plot in Fig.~\ref{fig:Anisotropy2D} a single ion magnetic anisotropy energy $-\bold{M}  \cdot\mu_0\bold{H}$ at $T=2$~K and $\mu_0H=3$~T, where $\bold{M}=\bold{M}(\bold{H},T)$. Only the Jahn-Teller effect is included in these calculations with no contribution from trigonal deformation. The applied external magnetic field vector $\bold{H}$ is represented using spherical coordinates. Such approach allows us to compare quantum $-\bold{M}^q\cdot\mu_0 \bold{H}$ (Fig.~\ref{fig:Anisotropy2D}.a) and classical $-\bold{M}^c \cdot\mu_0 \bold{H}$ (Fig.~\ref{fig:Anisotropy2D}.b) results, where ${M}^q$ corresponds to the magnetization obtained from CFM simulations, whereas ${M}^c$ describes the classical magnetization computed using the sLLG equation. In Fig.~\ref{fig:Anisotropy2D}, we observe very similar qualitative results obtained from both approaches. The easy magnetic axes are along three different Jahn-Teller directions $\bold{e}_{JT}$. The satisfactory fit of classical results to quantum ones is obtained by taking $K_\textrm{JT}=0.75$~meV.
 
 \begin{figure}[hbt]
  \centering
  \includegraphics[width=7.6 cm]{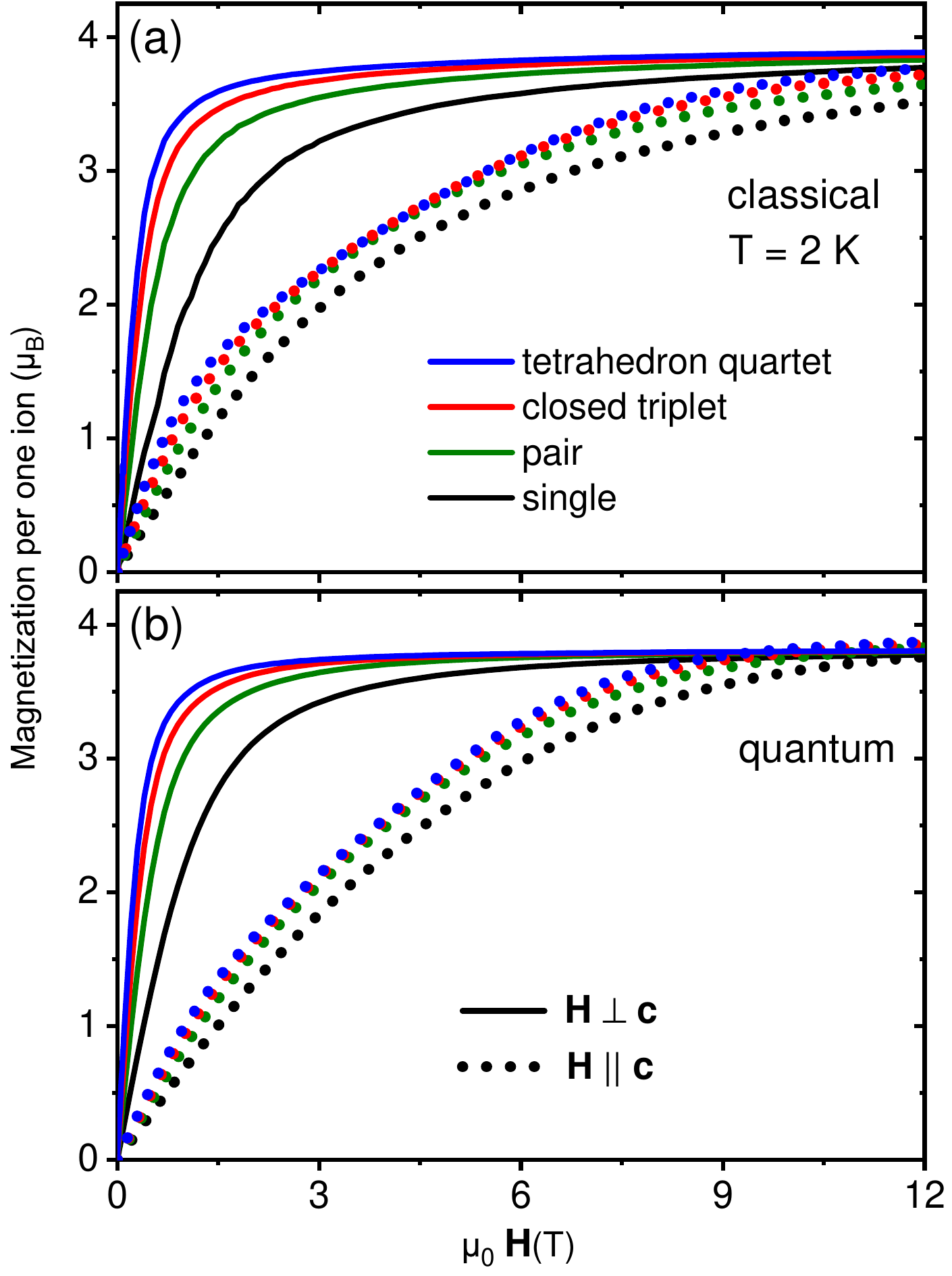}
  \caption{(Color online) Magnetization per one ion as a function of the magnetic field of different magnetic clusters using (a) classical and (b) quantum approaches at $T=2$~K. Solid and dotted lines denote magnetic easy ($\bold{H} \perp \bold{c}$) and hard axis ($\bold{H}$ $||$ $\bold{c}$) respectively. The magnetization per ion increases with the number of ions $N$ in a given cluster due to the presence of ferromagnetic superexchange coupling between Mn$^{3+}$ ions.}
  \label{fig:MH_anizo}
\end{figure}
 
\begin{figure}[hbt]
  \centering
 \includegraphics[width=7.6 cm]{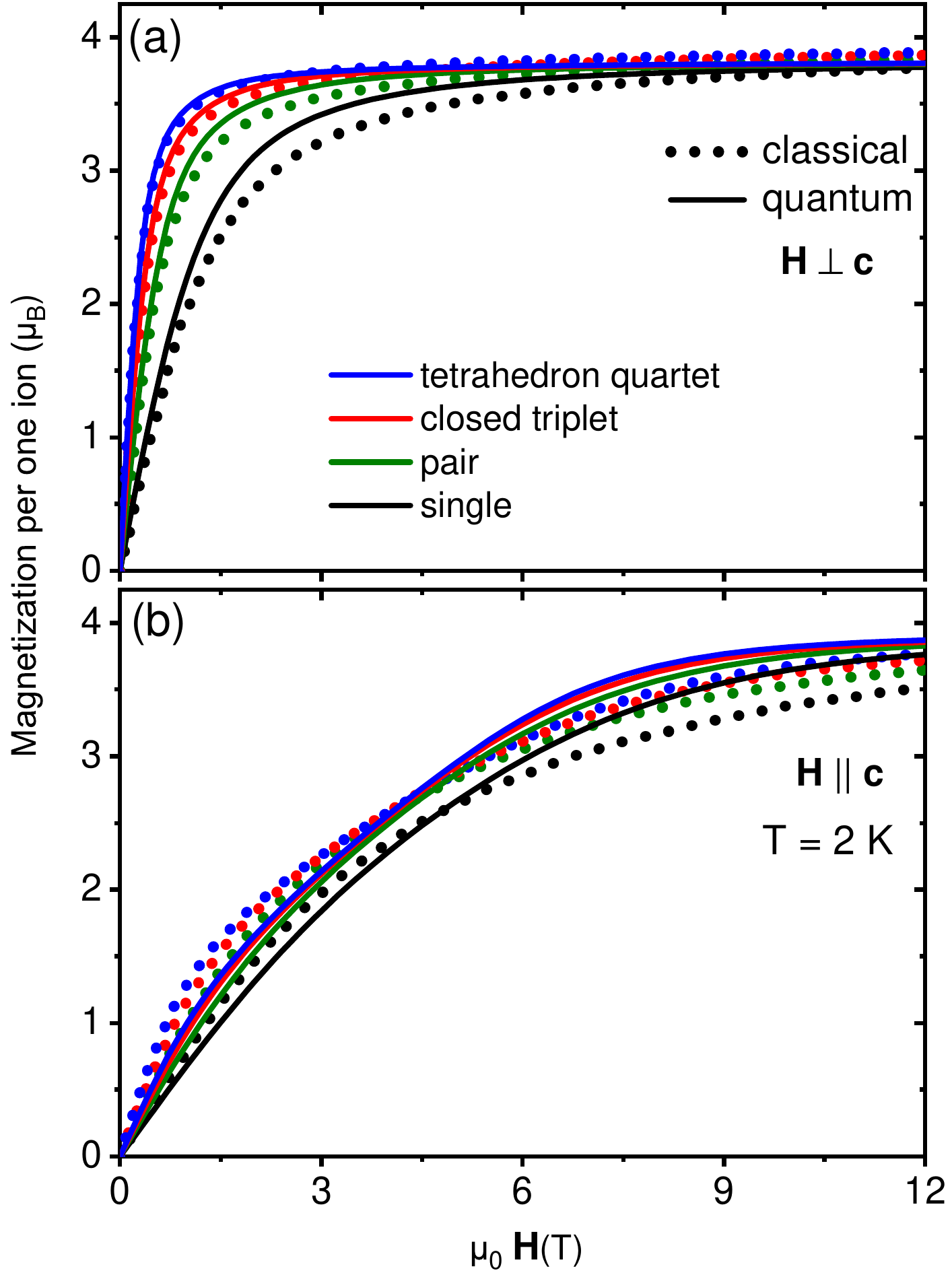}
  \caption{(Color online) Comparison of magnetization per one ion as a function of the magnetic field of different magnetic clusters at $T=2$~K with field applied along the (a) magnetic easy ($\bold{H} \perp \bold{c}$) and (b) hard axis ($\bold{H}$ $||$ $\bold{c}$). Solid and dotted lines denote quantum  and classical simulations respectively .}
  \label{fig:MH_comparison}
\end{figure}

Next, in Fig.~\ref{fig:MH_anizo} we plot the magnetization per one ion of different magnetic clusters coupled by the ferromagnetic superexchange $J^q=0.5$~meV as a function of the magnetic field $\bold{H}$ at $T=2$~K. As we can see from Fig.~\ref{fig:MH_anizo}, the magnetization per ion increases with the number of ions $N$ in a given cluster. This increase in magnetization with $N$ is expected since a ferromagnetic superexchange coupling exists between the Mn$^{3+}$ ions. 
However, the most prominent feature is the presence of a significant magnetic anisotropy, with the easy axis perpendicular to the $\bold{c}$ axis of GaN ($\bold{H} \perp \bold{c}$), and the hard one parallel to $\bold{c}$ ($\bold{H}$ $||$ $\bold{c}$). This magnetic anisotropy is mostly controlled by trigonal deformation \cite{gosk2005magnetic, stefanowicz2010structural, sztenkiel2016stretching, sztenkiel2020crystal}.
In experimental conditions, such deformation can be modified by the external electric field through the inverse piezoelectromagnetic effect or by stretching the material epitaxially \cite{sztenkiel2016stretching}.
By setting $K_{TR}=-1.24$~meV in classical simulations, we reproduce quantum curves both qualitatively and quantitatively, as shown in Fig.~\ref{fig:MH_comparison}. Having adjusted classical parameters $K_{TR}$ and $K_\textrm{JT}$ at $T=2$~K, we can perform comparative studies of magnetization obtained from the quantum and the classical approach as a function of $T$.

\begin{figure*}[htb]
  \centering
 \includegraphics[width=12.0 cm]{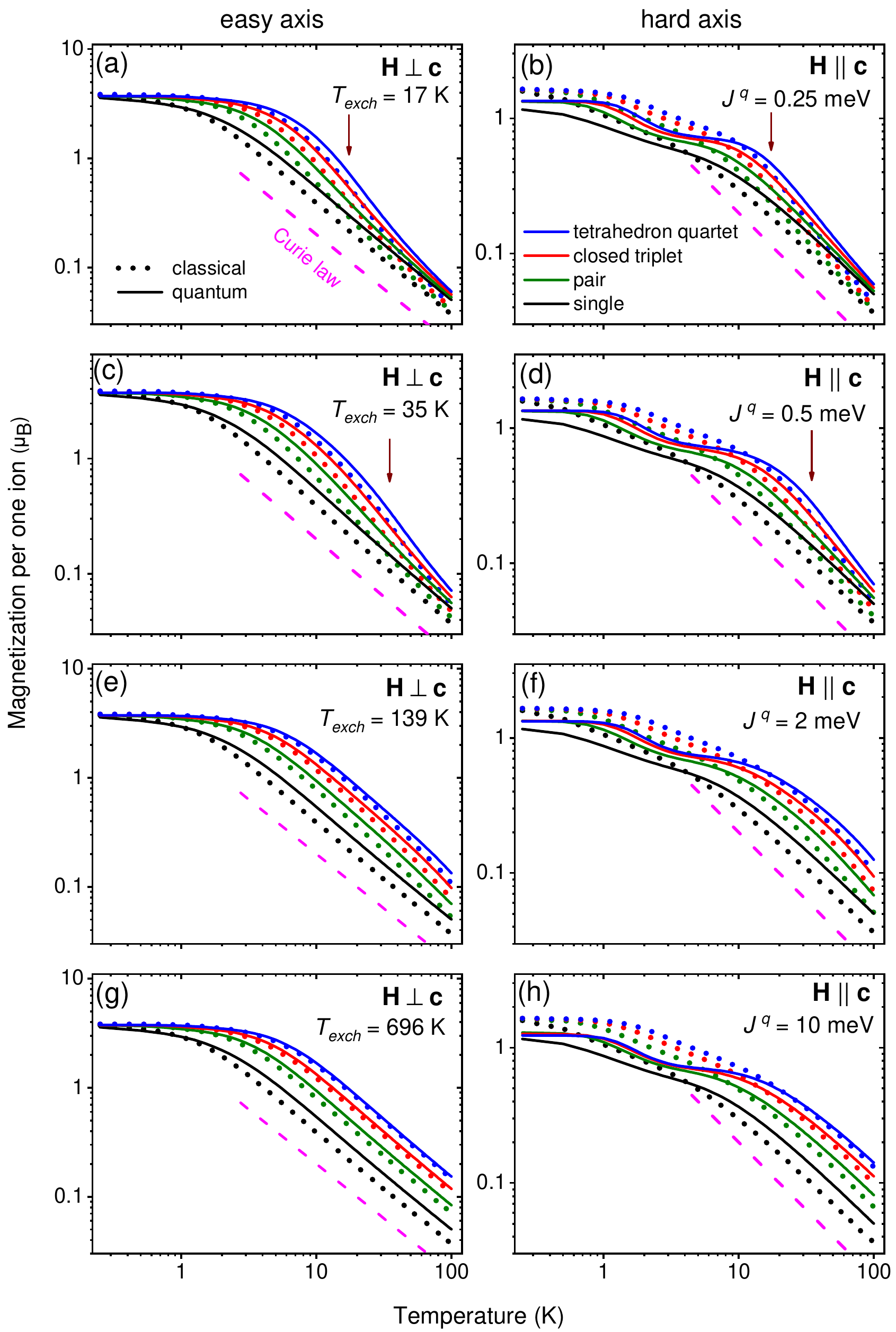}
\caption{(Color online) Magnetization versus temperature of different magnetic
clusters in an external magnetic field of $\mu_0H=1$~T applied along the magnetic easy ($\bold{H} \perp \bold{c}$) and hard axis ($\bold{H}$ $||$ $\bold{c}$). Solid and dotted lines denote quantum and classical simulations respectively. The super-exchange coupling $J^q$ is 0.25~meV (a and b), 0.5~meV (c and d) , 2~meV (e and f) and 10~meV (g and h). The dashed magenta lines indicate the slope of the Curie law. Arrows in panels (a)-(d) indicate the value of $T_{exch} = J^q |\bold{\hat{S}}_i| \cdot |\bold{\hat{S}}_j| / k_B$, that is the temperature at which the magnitude of exchange energy equals the thermal energy. In panels (e)-(h) $T_{exch} > 100$~K.}
 \label{fig:MT}
\end{figure*}


\subsection{Temperature-dependent magnetization}
\label{sec:MT}


Studies on the magnetization of various clusters as a function of temperature (up to 100~K) are shown in Fig.~\ref{fig:MT}. Since the calculations are computationally demanding, we restrict the upper limit of simulation temperatures to 100~K. 
Both in classical and quantum simulations, a constant magnetic field of $\mu_0H=1$~T is implemented. We present results for $\bold{H}$ applied both along the magnetic easy (Fig.~\ref{fig:MT}.a, c, e, g) and the hard axis (Fig.~\ref{fig:MT}.b, d, f, h). The magnetization is calculated for four different values of $J^q$, namely 0.25, 0.5, 2, 10~meV. The $M(T)$ curves are presented in a double logarithmic scale. As shown in Fig.~\ref{fig:MT}, at the lowest temperatures $T \lesssim 1$~K, the magnetization saturates. This effect is more pronounced for larger clusters. At elevated temperatures, $T \gtrsim 20$~K one would expect that ${M}(T)$ follows the Curie law (represented by dashed magenta lines in Fig.~\ref{fig:MT}). However, the $1/T$ dependency is obeyed only by singles, and larger clusters with the magnitude of exchange energy $J^q |\bold{\hat{S}}_i| \cdot |\bold{\hat{S}}_j|$ greater than the thermal energy $k_B T$  ($0.1 \leqslant T \leqslant 100$~K), as shown in Fig.~\ref{fig:MT} e-h. On the contrary, when $J^q |\bold{\hat{S}}_i| \cdot |\bold{\hat{S}}_j|<k_B T$, the individual spins creating a given cluster are partially decoupled. Then the decrease in magnetization ${M}(T)$ with increasing $T$ is steeper for such clusters than for the magnetization of single ion and ${M}(T)$ expected from Curie law. At the highest temperatures investigated here, the magnetization of larger clusters with low $J^q$ approaches the single ion results. This situation is exemplified in Fig.~\ref{fig:MT} a-d, with $J^q=0.25$ and $0.5$~meV. The detailed analysis of this effect, with an additional comparison with experimental data, is also reported in Ref.~\onlinecite{sztenkiel2020crystal}. 

Interestingly, we observe a small valley in the quantum $M^q(T)$ curves centered at $T \approx 3$~K when the magnetic field is applied along the hard axis ( $\bold{H}$ $||$ $\bold{c}$), as shown in Fig.~\ref{fig:MT} b, d, f, h. The magnitude of this dip increases with the cluster's size, and its center slightly shifts to higher temperatures with $N$. It was only possible to partially reproduce this effect using classical simulations. Nevertheless, this phenomenon awaits its experimental verification.

\begin{figure*}[htb]
  \centering
 \includegraphics[width=12.0 cm]{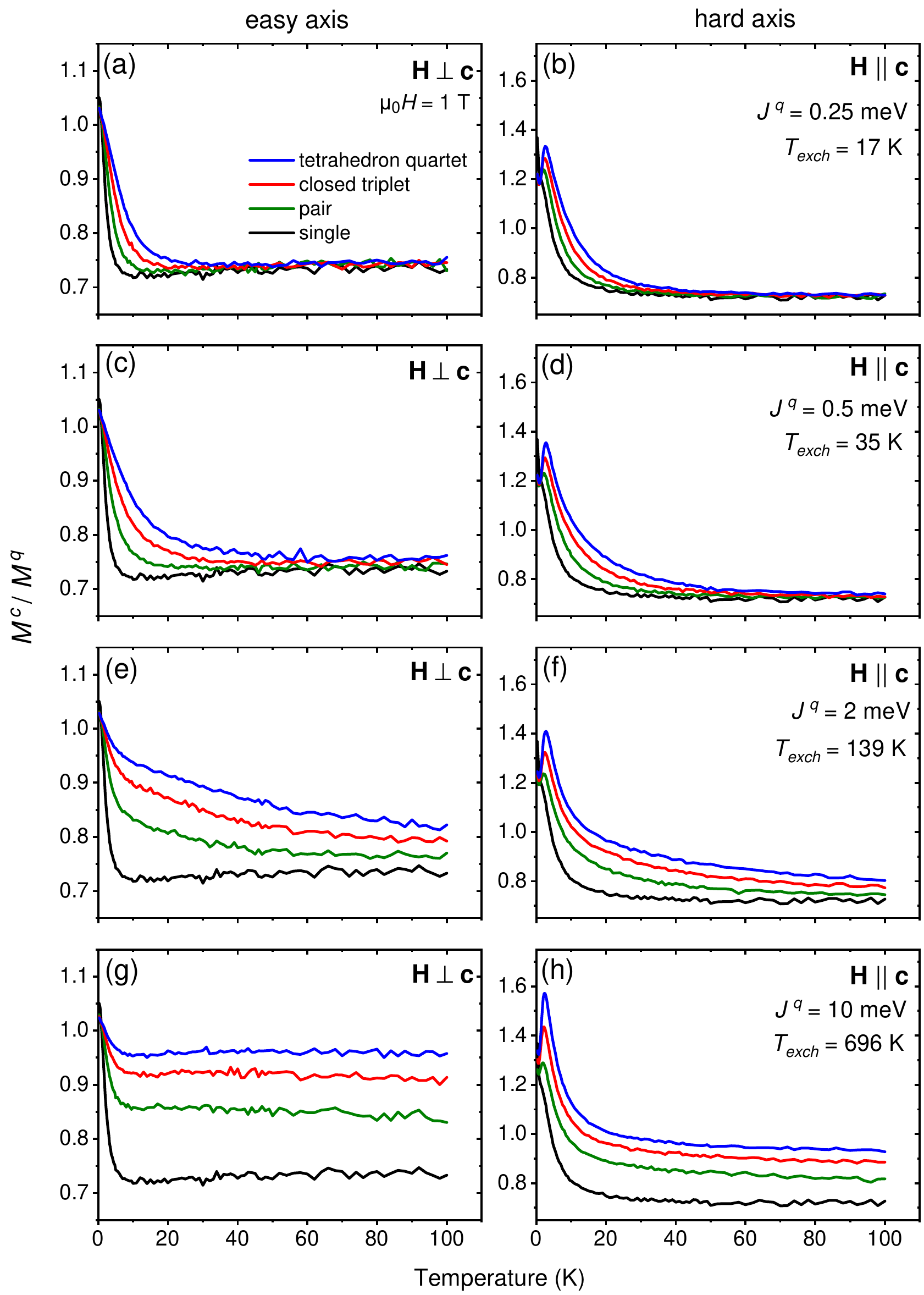}
        \caption{(Color online) Comparison of finite temperature magnetization studies using classical (${M}^c$) and quantum (${M}^q$) simulations. An external field of $\mu_0H=1$~T is applied along magnetic easy ($\bold{H} \perp \bold{c}$) and hard axis ($\bold{H}$ $||$ $\bold{c}$) with $J^q$= 0.25~meV (a and b), 0.5~meV (c and d) , 2~meV (e and f) and 10~meV (g and h). The $M^c / M^q$ ratios were calculated using the data presented in Fig.~\ref{fig:MT}. $T_{exch} = J^q |\bold{\hat{S}}_i| \cdot |\bold{\hat{S}}_j| / k_B$ describes the temperature at which the magnitude of exchange energy equals the thermal energy.}
        \label{fig:MLLG_MCF}
\end{figure*}

\subsection{Comparative analysis of temperature-dependent magnetization obtained using classical and quantum simulations}
\label{Sec:Comparison}



In the previous section, we can observe a discrepancy between results of classical and quantum simulations presented in $\bold{M}(T)$ graphs. In order to shed light on this disparity, we calculate the ratio of classical magnetization to the quantum one $M^{c}/M^{q}$ for different values of $J^q$ at $\mu_0 H = 1$~T, as shown in Fig.~\ref{fig:MLLG_MCF}. Ideally, this ratio should be exactly one in the whole temperature range. However, the presented data indicate that $M^{c}/M^{q}$ depends on the temperature, the value of super-exchange coupling, and the size of the cluster $N$. First, we focus on results obtained for the single non-interacting ion (black lines in Fig.~\ref{fig:MLLG_MCF}). 
For the magnetic easy axis ($\bold{H} \perp \bold{c}$), the ratio $M^{c}/M^{q}$ approaches unity at very low temperatures $T\lesssim2$~K (please note that we adjust the classical parameters at $T = 2$~K, as shown in Fig.~\ref{fig:MH_comparison}) . For $\bold{H}$ $||$ $\bold{c}$, this ratio is slightly higher that one $M^{c}/M^{q} \simeq 1.2$ at $T\lesssim2$~K, due to the fact that it was not possible to perfectly reproduce the shape of $M^{q}(\bold{H})$ using classical simulations, as seen in Fig.~\ref{fig:MH_comparison}. As the temperature increases, $M^{c}/M^{q}$ ratio decreases rather fast both for the easy and hard magnetization, and saturates at the value of $M^{c}/M^{q} \simeq  0.75$ for $T \gtrsim 10$~K. Surprisingly, at very high temperatures, where quantum and classical results should merge, the $M^{c}/M^{q}$ ratio is low and significantly below unity. The reason for this is explained in section \ref{Sec:Analysis}. 

Now we turn to the data obtained for larger clusters and small values of exchange coupling $J^q=0.25$ and $0.5$~meV. We find here that the $M^{c}/M^{q}$ ratio clearly increases with the cluster size at low temperatures, $T \leqslant 30$~K, as shown in Fig.~\ref{fig:MLLG_MCF} a-d. However $M^{c}/M^{q}$ still decreases with $T$. At high temperatures, where $J^q |\bold{\hat{S}}_i| \cdot |\bold{\hat{S}}_j| < k_B T$, results for larger clusters with $N \geq 2$ approach the single ion value, as expected. We observe an improvement in the comparison of the results of classical and quantum simulations with the increase of the exchange coupling $J^q$ to 2~meV. Finally, as shown in Fig.~\ref{fig:MLLG_MCF} g-h, $M^{c}/M^{q}$ is close to unity for the tetrahedron quartet with $J^q = 10$~meV in almost whole studied temperature range.


To sum up, we see that the $M^{c}/M^{q}$ ratio depends strongly on the value of super-exchange coupling $J$ and the size of the cluster $N$. The comparison between results of classical and quantum simulations improves with increasing magnitude of $J^q$. It seems, that for $J^q |\bold{\hat{S}}_i| \cdot |\bold{\hat{S}}_j| < k_B T$ the $M^{c}/M^{q}$ ratio is rather low. However, when $J^q |\bold{\hat{S}}_i| \cdot |\bold{\hat{S}}_j| \gg k_B T$, the $M^{c}/M^{q}$ ratio for $N \geq 2$ approaches unity, and it is close to one for $N=4$ in almost whole studied temperature range.  

\subsection{Analysis}
\label{Sec:Analysis}

\begin{figure*}[hbt]
 \centering
 \includegraphics[width=12 cm]{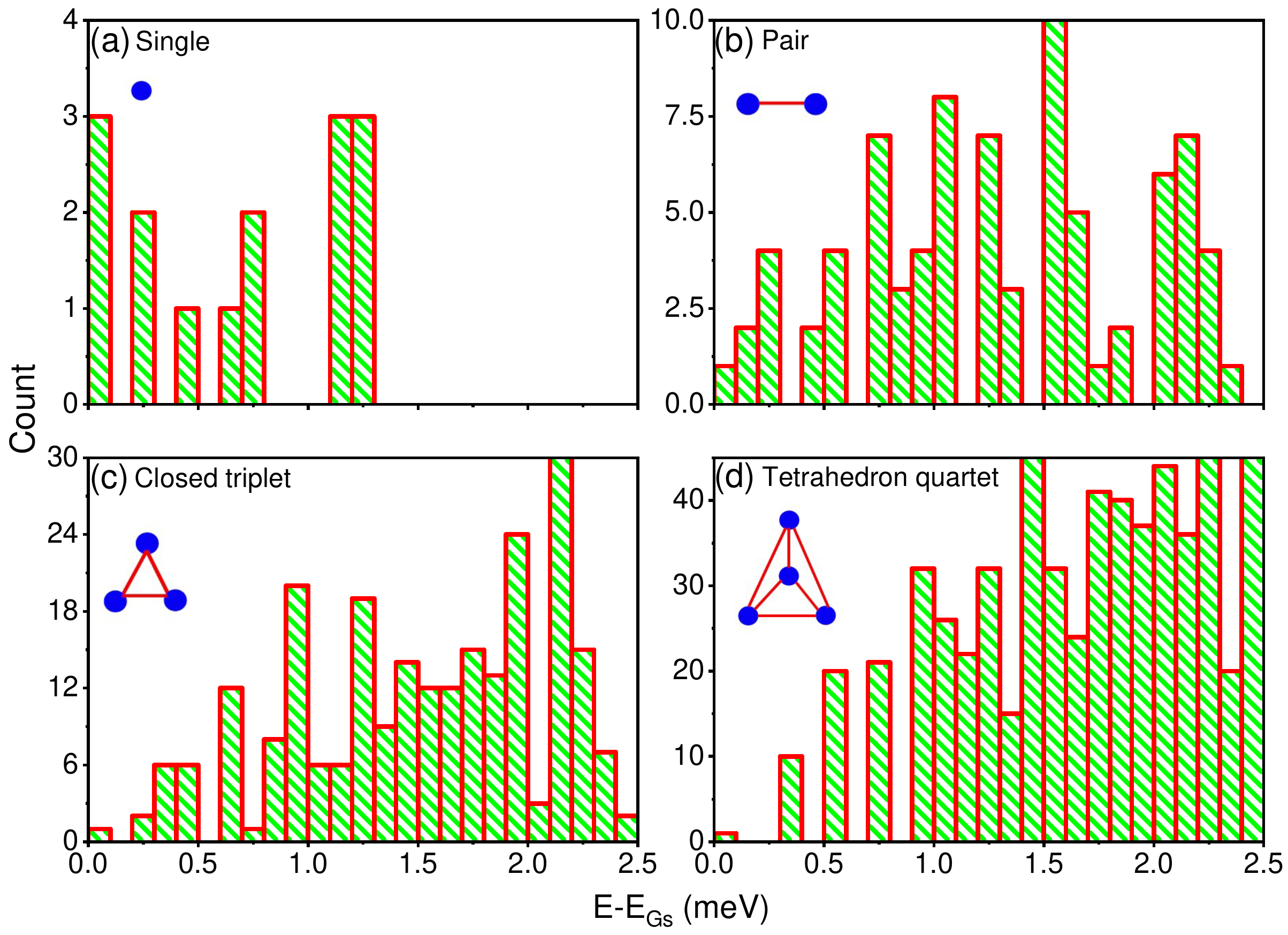}
\caption{(Color online) Available number of states contained in the energy interval ($E$, $E+\Delta E$), with $\Delta E = 0.1$~meV for (a) singlet (b) pair (c) closed triplet (d) tetrahedron quartet with $J^q$ = 0.5~meV and $\mu_0 H = 1$~T. The magnetic field is applied along the hard axis $\bold{H}$ $||$ $\bold{c}$. The energy is presented relative to the energy of the ground state $E_{GS}$.}
\label{fig:energylevels}
\end{figure*}

In Fig.~\ref{fig:MLLG_MCF}, we observed a significant discrepancy between the results of quantum and classical simulations as a function of the temperature. As stated previously, the above disparity stems from the classical nature of the sLLG model \cite{Kuzmin:2005_PRL,evans2015quantitative}. Generally, one can compute $M(T)$ using the classical approach and reproduce rather well the Curie temperature $T_\mathrm{C}$, but not the shape of experimental curves \cite{evans2015quantitative}. Even in the paramagnetic case studied here, the classical treatment deviates significantly from results obtained using the quantum treatment. Several methods have been proposed in the past to resolve this issue \cite{Kuzmin:2005_PRL,evans2015quantitative,Bergqvist:2018_PRM, Woo:2015_PRB}. For example, Kuz’min introduced a simple phenomenological equation to quantify the shape of spontaneous $M(T)$ \cite{Kuzmin:2005_PRL}. Other approaches proceed by implementing temperature rescaling \cite{evans2015quantitative} or by incorporating quantum Bose-Einstein statistics instead of classical Boltzmann statistics \cite{Bergqvist:2018_PRM}. The rescaling approach allows to modify the classical thermodynamic quantities toward the quantum mechanical solution and include in the simulations the underlying quantum effects in $ad$  $hoc$, and approximate way \cite{Korman:2010_PRB}. However, these methods are primarily applicable to the bulk material with a well defined critical temperature $T_C$.

To have a better insight into why the $M^{c}/M^{q}$ ratio is small for low values of $J^q$, we calculated a quasi density of states (qDOS), which describes the number of quantum states available within an energy range from $E$ to $E+\Delta E$. The input data come from CFM simulations, and the results are shown in Fig.~\ref{fig:energylevels}. A simple explanation of the disparity between classical and quantum simulation results refers to the concept of enhanced thermal spin fluctuations in the classical limit owing to the effect of quantum mechanical “stiffness” \cite{evans2015quantitative}. The physical interpretation is that in quantum mechanics, only specific discrete energy levels exist, as shown in Fig.~\ref{fig:energylevels} a, for the case of a single non-interacting Mn$^{3+}$ ion. When the temperature $T$ is sufficiently low, the thermal energy $k_{B}T$ becomes comparable with the energy difference between the discrete quantum levels. The system populates mostly the ground state with a low probability of occupying higher excited states. This dramatically reduces the thermal fluctuations of the quantum system. On the contrary, in the classical case, a continuous energy spectrum is observed, and the allowed spin fluctuations are enhanced compared to the quantum case. Hence, the classical approach fails in describing the magnetization at low temperatures where $M^{c} < M^{q}$. On increasing temperature, the probability of occupying higher excited states increases and the quantum mechanics gradually merges with the classical one, bringing $M^{c}$ closer to $M^{q}$. Therefore, one might expect that quantum simulations will give enhanced results than the classical ones at lower temperatures, and the $M^{c}/M^{q}$ ratio should increase with $T$ when the quantum system approaches the classical limit.  However, a significant departure from this expected characteristic is presented in Fig.~\ref{fig:MLLG_MCF}. For low values of $J^q$, the $M^{c}/M^{q}$ ratio acquires maximum (not minimum) values at lower temperatures, and it decreases with $T$ (instead of increasing towards unity). Finally, $M^{c}/M^{q}$ saturates at low level for $T \gtrsim 50$~K. 

The figure~\ref{fig:energylevels} also illustrates a substantial variation of qDOS for different cluster types. It is of a discrete type for the singlet,  but acquires a quasi-continuous form for the pair, closed triplet and tetrahedron quartet. This gradual transition from the discrete to the quasi-continuous form of qDOS as the cluster size increases is consistent with the data presented in Fig.~\ref{fig:MLLG_MCF} c,d. The correspondence between classical and quantum results is the best at $T<20$~K for tetrahedron quartet (as compared with other clusters with $N<4$), that is for the largest cluster whose quantum level structure, depicted in Fig.~\ref{fig:energylevels} d, most closely resembles the classical case. The conclusion is that, it is possible to only partially explain the inconsistency between classical and quantum results presented in temperature-dependent magnetization curves by referring to the concept of quantum mechanical “stiffness”.

Instead, one can invoke simple analytic expressions for magnetic susceptibility in order to explain the observed differences between $M^{c}$ and $M^{q}$. It is well known that in the isotropic case, the magnetization of a single ion can be calculated analytically, using both quantum mechanical and classical considerations, by referring to the Brillouin and Langevin functions, respectively. When $\mu_B \mu_0 H / k_B T$ is small (low to moderate $\mu_0 H$, high $T$), it is possible to derive a quantum $\chi^q = \frac{\mu_0 g^2 \mu_B^2 S(S+1)}{3 k_B T}$ and a classical $\chi^c = \frac{\mu_0 g^2 \mu_B^2 S^2}{3 k_B T}$ approximations of magnetic susceptibilities (the Curie law). It is of interest to compare the two expressions. Clearly, the ratio $\frac{\chi^c}{\chi^q} = \frac{S}{S+1}$ depends on the value of spin $S$, and both approaches merge for $S \rightarrow \infty $.  In our opinion, the presented above relation explains rather well the temperature-dependent data depicted in Fig.~\ref{fig:MLLG_MCF}. For single non interacting ions, $M^{c}/M^{q}$ saturates for high temperatures at $M^{c}/M^{q} \simeq  0.75$. This saturation value agrees reasonably well with the analytical level $\chi^c / \chi^q = 2/3$ obtained for $S=2$. The observed small discrepancy between these two magnitudes originates from the presence of the magnetic anisotropy in our numerical calculations, whereas $\chi^c / \chi^q$ was obtained for an isotropic case. Similarly, for larger clusters and a small value of exchange coupling $J^q=0.25$ and $0.5$~meV (Fig.~\ref{fig:MLLG_MCF} a-d), the individual spins are practically decoupled at $T \gtrsim 50$~K, and $M^{c}/M^{q}$ for $N \geqslant 2$ approaches the single ion value. Interestingly, the increase of temperature does not lead to the merging of classical and quantum results as $M^{c}/M^{q}$ have to stay close to the fundamental single ion limit defined by $\chi^c / \chi^q = 2/3$ with $S=2$. However, in the case when the exchange energy is very strong $J^q =10$~meV, that is $J^q |\bold{\hat{S}}_i| \cdot |\bold{\hat{S}}_j| \gg k_B T$, we observe quite different characteristics (Fig.~\ref{fig:MLLG_MCF} g-h). At high temperatures, the $M^{c}/M^{q}$ ratio increases with $N$, and approaches unity for $N=4$. For $J^q =10$~meV, the Mn$^{3+}$ ions are strongly bound in the whole temperature range investigated here, $0.25 \geqslant T \geqslant 100$~K, behaving as an effective single ion with the total spin of $S_T = N S$. Then $M^{c}/M^{q}$ ratio at high $T$ follows roughly a simple relation $\chi^c / \chi^q =NS /(NS+1)$, which is equal to 4/5, 6/7 and 8/9 for the pair, triplet and quartet respectively. It is consistent with the simple fact, that the classical and quantum simulations merge for $S \rightarrow \infty$. However, in this paper, depending on the relative strength between the thermal and the exchange energy, we deal with uncoupled ions characterized by $S=2$ or strongly bound clusters with large total spin value $S_T = N S$.

The analysis presented above explains why using the classical approach, it is very hard to properly model the shape of the experimental magnetization curves of different ferromagnetic materials as a function of temperature. The problem is the incorrect form of the classical $M(T)$ curve when compared to experiment \cite{evans2015quantitative, Bergqvist:2018_PRM, Korman:2011_PRB}. At low temperatures, well below $T_C$, the magnetic ions are strongly bound creating percolating magnetic system with a high total spin $S_T$. In this regime, the classical approach can give correct and satisfactory results. However, by increasing $T$, but still below $T_C$, spins become weakly coupled (a large effective cluster gradually breaks down into small pieces with low $S$), and then the classical approximation underestimates the actual value of the magnetization. 

These conclusions are substantiated by the results shown in Fig.~\ref{fig:MLLG_MCF} e-f. For $J^q =2$~meV the magnitude of the exchange energy is still slightly greater than the thermal energy in the entire studied temperature range, that is $J^q |\bold{\hat{S}}_i| \cdot |\bold{\hat{S}}_j| > k_B T$. However $M^{c}/M^{q}$ for larger clusters decreases with increasing $T$ and gradually approaches the single ion limit. Only in the limit of very strong coupling with $J^q =10$~meV, where $J^q |\bold{\hat{S}}_i| \cdot |\bold{\hat{S}}_j| > 6 k_B T$, the given cluster acts as an effective single-ion spin with $S_T=NS$, as shown in Fig.~\ref{fig:MLLG_MCF} g-h.

\section{CONCLUSIONS}

We have performed a comparative study of magnetization $\bold{M}(\bold{H},T)$ of Mn$^{3+}$ ions ($S=2$, $L=2$) in GaN using classical and quantum mechanical approaches. We have investigated small magnetic clusters consisting of isolated ions, pairs, triplets, and quartets of Mn$^{3+}$ ions coupled by the ferromagnetic super-exchange interaction. We have found significant differences between the results of both approaches (classical and quantum crystal field model). Especially, we found that the ratio of classical magnetization to the quantum one $M^{c}/M^{q}$ strongly varies with temperature and it depends on the choice of the exchange coupling constant $J^q$ and the size of the cluster $N$. The $M^{c}/M^{q}$ ratio improves with increasing of $J^q$ and approaches unity in the strong coupling limit $J^q |\bold{\hat{S}}_i| \cdot |\bold{\hat{S}}_j| \gg k_B T$ and for large value of $N$. Our results show that it is very complex task, to reproduce the complicated quantum crystal field model CFM using a simple classical approximation. We have found that the key factor is the magnitude of the effective spin of the system, which depends on the relative strength between the thermal and the exchange energy. For weak coupling between the spins $J^q |\bold{\hat{S}}_i| \cdot |\bold{\hat{S}}_j| < k_B T$, the investigated clusters consist of practically uncoupled spins characterized by $S=2$. Then classical approximation underestimates the actual value of magnetization. Contrary, in the strong coupling regime, characterized by $J^q |\bold{\hat{S}}_i| \cdot |\bold{\hat{S}}_j| \gg k_B T$, we deal with one magnetic system characterized by high total spin value $S_T = NS$. In this limit, $S_T$ increases with the number of ions $N$ in the cluster, what improves the comparison between $M^{c}$ and $M^{q}$. These conclusions are consistent with the simple fact, that the classical and quantum simulations differ for small value of $S$ and both treatments merge for $S \rightarrow \infty$.

\section*{Acknowledgments}

We would like to thank M. Sawicki for proofreading of the manuscript and valuable suggestions. The work is supported by the National Science Centre (Poland) through project OPUS 2018/31/B/ST3/03438. This research was carried out with the support of the Interdisciplinary Centre for Mathematical and Computational Modelling (ICM) University of Warsaw under grant no GB77-6

\medskip

\end{document}